\documentclass{aa}
\usepackage{graphicx,psfig}

\makeatletter
\let\@internalcite\cite
\def\cite{\@ifstar{\citeyear}{\citefull}}
\def\citefull{\def\astroncite##1##2{##1 ##2}\@internalcite}
\def\citeyear{\def\astroncite##1##2{##2}\@internalcite}
\def\citeau{\def\astroncite##1##2{##1}\@internalcite}
\def\citen{\def\astroncite##1##2{##1 (##2)}\@internalcite}
\def\possesivcite{\def\astroncite##1##2{##1's (##2)}\@internalcite}
\def\@citex[#1]#2{\if@filesw\immediate\write\@auxout{\string\citation{#2}}\fi
  \def\@citea{}\@cite{\@for\@citeb:=#2\do
    {\@citea\def\@citea{; }\@ifundefined
       {b@\@citeb}{{\bf ?}\@warning
       {Citation `\@citeb' on page \thepage \space undefined}}%
{\csname b@\@citeb\endcsname}}}{#1}}
\def\@cite#1#2{#1\if@tempswa , #2\fi}
\def\@biblabel#1{}
\makeatother

\newcommand{\beqa}{\begin{eqnarray*}}
\newcommand{\eeqa}{\end{eqnarray*}}
\newcommand{\beqan}{\begin{eqnarray}}
\newcommand{\eeqan}[1]{\label{#1}\end{eqnarray}}
\newcommand{\beq}{\begin{equation}}
\newcommand{\eeq}{\end{equation}}

\begin{document}

\thesaurus{
02.08.1; 
08.09.3, 
08.18.1. 
}

\title{The Nature of the Lithium Rich Giants}
\subtitle{Mixing episodes on the RGB and early-AGB}

\author{Corinne Charbonnel \inst{1},  Suchitra C. Balachandran \inst{2}}

\offprints{Corinne Charbonnel ; corinne@obs-mip.fr} 

\institute{
Laboratoire d'Astrophysique de l'Observatoire Midi-Pyr\'en\'ees, CNRS UMR 5572, 
14, Av. E.Belin, 31400 Toulouse, France 
\and 
Department of Astronomy, University of Maryland, College Park, MD 20742,
USA}

\date{Received 2000 February 16 / Accepted 2000 May 12}

\maketitle
\markboth{C. Charbonnel and S.Balachandran : 
The nature of the lithium rich giants}{}

\begin{abstract}
We present a critical analysis of the nature of the so-called Li-rich
RGB stars.
For a majority of the stars, we have used Hipparcos parallaxes to 
determine masses and evolutionary states by comparing their position 
on the Hertzsprung-Russell diagram with theoretical evolutionary tracks.
Among the twenty Li-rich giants whose location on the HR diagram we were able
to determine precisely, five appear to be Li-rich because they have not
completed
the standard first dredge-up dilution, and three have abundances compatible
with the maximum allowed by standard dilution.  Thus, these should 
be re-classified as Li-normal.  
For the remaining stars, the high Li abundance must be a
result of fresh synthesis of this fragile element.

We identify two distinct episodes of Li production which occur
in advanced evolutionary phases depending upon the mass of the star.  
Low-mass RGB stars, which later
undergo the helium flash, produce Li at the phase
referred to as the bump in the luminosity function.  At this evolutionary
phase, the outwardly-moving hydrogen shell burns through the
mean molecular weight discontinuity created by the first dredge-up.
Any extra-mixing process can now easily connect the
$^3$He-rich envelope material to the outer regions of the 
hydrogen-burning shell, enabling Li production by
the Cameron \& Fowler (1971) process.
While very high Li abundances are then reached,
this Li-rich phase is extremely short lived because once the mixing extends
deep enough to lower the carbon isotopic ratio below the standard dilution
value, the freshly synthesized Li is quickly destroyed.

In intermediate-mass stars, the mean molecular weight gradient due
to the first dredge-up is not erased until after the star has 
begun to burn helium in its core.  The Li-rich phase in these
stars occurs when the convective envelope deepens at the base of
the AGB, permitting extra-mixing to play an effective role. 
Li production ceases when a strong mean molecular weight gradient is 
built up between the deepening convective envelope and the 
shell of nuclear burning that surrounds the inert CO core.
This episode is also very short lived.
Low-mass stars may undergo additional mixing at this phase.

The compiled data provide constraints on the time scales for extra mixing
and some insight on processes suggested in the literature.
However, our results do not suggest any specific trigger mechanism.
Since the Li-rich phases are extremely short, enrichment of the Li
content of the ISM as a result of these episodes is negligible.

\keywords{Li dip; hydrodynamics; turbulence; 
Stars: interiors, rotation, abundances}

\end{abstract}

\section{The so-called Li-rich RGB stars}
During the first dredge-up, the lithium abundance 
at the surface of a red giant star decreases due to dilution of the external 
convective stellar layers with the lithium-free region in the interior.  
Depending on the stellar mass and metallicity, 
the surface lithium abundance decreases with respect to its value at the 
end of the main sequence by a factor that varies between $\sim$ 
30 and 60.
The post dredge-up lithium abundance also depends on the surface depletion
of this fragile element during the pre-main sequence and main sequence
phases; in Population I stars this is known to be important at
masses lower than $\sim$
1.2 M$_{\odot}$ and in stars originating from the Li dip 
(see volume edited by \cite{Crane94} for reviews). 
Starting from the present interstellar medium abundance of log N(Li)$\simeq$3.3
(where log N(Li) = log[n(Li)/n(H)]+12), 
one thus expects a post-dilution value lower than about 
1.8 to 1.5 for Pop I 
stars and, indeed, most G-K giants fall below this upper limit 
(\cite{Lambertetal80}, \cite{Brownetal89}, \cite{Mallik99}).

However, about 1$\%$ of G-K giants show unexpectedly strong lithium lines
(\cite{WS82}, \cite{Brownetal89}, \cite{GrattonDAntona89}, 
\cite{Pilachowskietal90}, \cite{Pallavicinietal90}, 
\cite{FekelBalachandran93}).  
Some of these Li-rich giants have abundances that are even higher than 
the present interstellar medium value 
(\cite{DeLaRezaDaSilva95}, \cite{Balachandranetal00}).
Various suggestions have been made to explain the Li-rich giant phenomenon. 
Some are related to external processes, like the contamination of the
external layers of the giant by the debris of nova ejecta or by the 
engulfing of a planet (\cite{Alexander67}, \cite{Brownetal89},
\cite{GrattonDAntona89}, \cite{SiessLivio99}). 
Other explanations explore internal processes, like the preservation of
the initial lithium content or fresh lithium production 
(\cite{FekelBalachandran93}, \cite{DeLaRezaetal96}, \cite{Castilhoetal99}, 
\cite{SackmannBoothroyd99}). 

The other notable disagreement betwen the prediction of abundances in
first ascent giants and observations is the carbon isotopic ratio.
It has been observed that the carbon isotopic ratio in evolved stars 
of open clusters with turnoff masses lower than about 2.2M$_{\odot}$ 
(\cite{Gilroy89}, \cite{GilroyBrown91}) and in field giants at various 
metallicities
(\cite{Snedenetal86}, \cite{Shetroneetal93}, \cite{Pilachowskietal97},
\cite{cbw:98}, \cite{Carrettaetal98}, \cite{Grattonetal2000}) 
is lower than the value predicted by the standard theory. 
Observations reveal that the carbon isotopic
ratio does not decrease below the standard model predictions 
until the mean molecular weight gradient produced 
by the first dredge up is erased by the outwardly-burning hydrogen shell.
This evolutionary phase is referred to as the bump in the
luminosity function on the HR diagram and corresponds to a temporary
decrease in the luminosity and a small increase in the effective temperature 
of the star when the chemical discontinuity is removed.
It has therefore been surmised that a non-standard mixing process, 
previously inhibited by the mean molecular weight barrier, begins to act 
at this phase and results in ``extra mixing" of the convective zone material 
with regions hot enough to convert $^{12}$C to $^{13}$C 
(Sweigart \& Mengel 1979; Charbonnel 1994, 1995; Charbonnel et al. 1998).

The nature of the mechanism which produces 
the drop of the $^{12}$C/$^{13}$C ratio remains uncertain,
though rotation-induced mixing is a probable candidate.  
We do know that (i) 
the extra-mixing is inhibited by molecular weight gradients because,
as our previous discussion showed, 
it has not been observed to occur before the star enters the bump,
(ii) it occurs in $\sim 96 \%$ of the low-mass
stars (\cite{CharbonnelDias98}) and (iii) it destroys part of the $^3$He 
produced on the main sequence (as suggested first by \cite{rbw84}; see
also \cite{cchar:95}, \cite{hog:95} and \cite{SackmannBoothroyd99}). 
It may be responsible for other chemical anomalies. 
For instance, a significant decrease in the surface lithium abundance 
is seen in Population II giants which are at the red-giant bump
(\cite{Pilachowskietal93}, \cite{cchar:95}, \cite{Grattonetal2000}). 
The continuous decline of the carbon
abundance along the RGB, and the presence in the atmosphere of material
processed by the ONeNa-cycle (see \cite{kra:94} and \cite{daco:98} for
reviews) seen in globular cluster giants may also result 
from a manifestation of the extra-mixing process (see \cite{weissDC00}
and references therein).

In this paper we draw a connection between the Li enhancement and 
the carbon isotopic ratio decline in red-giants and provide evidence 
for the hypothesis outlined in \S2.  For the first time 
it is shown that two phases of mixing occur in Population I stars 
depending upon the mass of the star;
mixing occurs at the luminosity bump for low-mass stars and in the early-AGB
phase before the completion of the second dredge-up in both
low and intermediate-mass stars. 
Both lead to short-lived Li-rich phases. 
We discuss the similarities between the two events.

\section{The Hypothesis}
We suggest that the 
Li-rich RGB stars are formed mainly as a precursor
to the deeper mixing process which produces anomalously low carbon isotopic
ratios.  

In low-mass stars (i.e., stars which ascend the RGB with a degenerate
He core), the fresh synthesis of lithium occurs at the start of the
red giant luminosity bump phase when the outwardly moving hydrogen-burning
shell burns through the mean molecular weight discontinuity.
With the help of an extra-mixing process, the $^3$He-rich 
envelope\footnote{During the first dredge-up, the deepening convective envelope 
is enriched in $^3$He when it engulfs the $^3$He 
peak built up while the star was on the main sequence.
In the standard models, $^3$He survives in the star until it is
injected into the ISM by stellar winds and planetary
nebulae ejection}
can now be connected to the hotter inner region in the vicinity of the 
hydrogen-burning shell.  Li production proceeds via the 
\citen{CameronFowler71} mechanism in which
$^3$He is burned to create $^7$Be via $^3$He($\alpha$, $\gamma$)$^7$Be.  
The $^7$Be quickly decays into $^7$Li via $^7$Be(e$^-$,$\nu$)$^7$Li.
For the freshly synthesized $^7$Li to survive, the $^7$Be must be 
transported rapidly to cooler regions before the decay occurs.
Depending on the mixing efficiency, which may vary from star to star,
the $^7$Li may or may not reach the stellar surface. 
In either case, 
once mixing extends deep enough to convert additional $^{12}$C into $^{13}$C 
(as is observed to occur in low-mass giants) 
the surface material is exposed to temperatures higher than $^{7}$Li can
withstand and the freshly synthesized $^{7}$Li is steadily destroyed.
Therefore the Li-rich phase is fleeting; a star will not 
retain its peak Li abundance once the $^{12}$C/$^{13}$C ratio dips below the 
standard value.

The evolutionary sequence of an intermediate-mass star differs in 
one key aspect, and this leads to Li production at a different epoch of 
its life.  Evolution up the RGB is rapid for intermediate-mass stars,
and helium burning in the non-degenerate core 
has already begun before the hydrogen-burning shell burns through the
mean molecular weight discontinuity caused by the first dredge-up.  
Therefore no extra mixing is facilitated on the short RGB phase of these stars,
as is confirmed from the observations of the 
carbon isotopic ratio which agree with standard model predictions 
(\cite{cc94}). 
When the mean molecular weight gradient is finally erased, 
the star is burning He in its core which is surrounded by the 
hydrogen-burning shell, 
the convective envelope is very shallow, and distant (both in 
terms of mass and radius) from the burning regions, so that extra-mixing
may not be effective. 
However, once the core He-burning
is exhausted and the star ascends the early-AGB, the convective envelope 
deepens in response to the contracting, slightly degenerate CO core 
with the helium- and hydrogen-burning shells at its outer edge.  
We suggest that it is at this phase that an extra-mixing process may be most 
effective in bridging the small 
gap between the base of the convective zone and the 
H burning shell, producing $^7$Li via the Cameron-Fowler (1971) process.
When the convective envelope reaches the external part of the hydrogen
burning shell, 
a strong gradient of molecular weight will again inhibit the extra-mixing.
We thus expect this phase to be very short and localized on the HR diagram.
Note that the base of the convective zone of a low-mass star also
comes into close proximity to the hydrogen-burning shell at this evolutionary
phase, and therefore these stars have a second opportunity to become
Li-rich if extra-mixing occurs.

In stars with initial mass higher than about 4M$_{\odot}$ the hydrogen-
burning shell is extinguished by the deepening convective envelope
and significant changes in the surface abundance are seen when the
H-He discontinuity is penetrated: the second dredge-up.  Accordingly, 
we do not expect to see $^7$Li-rich stars in the early-AGB phase at 
masses greater than about 4 M$_{\odot}$.  However, we recall that it 
is in stars between 4M$_{\odot}$ and
8M$_{\odot}$ that hot-bottom burning leading to $^7$Li production
has been both observed and shown to be theoretically feasible at a later 
evolutionary phase (\cite{SmithLambert89}, \cite{SmithLambert90}, 
\cite{Plezetal93}, \cite{Smithetal95}, \cite{SackmannBoothroyd92},
\cite{ForestiniCharbonnel97}).  

While some clues suggest that the deep mixing may be related to
rotation, the physical description of the process is far from complete,
and observational evidence is inconclusive up to now. 
We endorse no specific mechanism in this study.  
We have no clues on whether the trigger for the extra-mixing is 
the same in the low- and intermediate-mass stars.
Rather, our hypothesis is based on a theoretical understanding of the
structure of the red-giant star, on the results of the $^{12}$C/$^{13}$C
ratio studies referenced above and on our own analysis of
the properties of Li-rich giants.
The novel aspect of our study is that we are able to substantiate 
our claim with a critical analysis of Li-rich giant data from the literature. 
In particular, we have used Hipparcos parallaxes to determine 
the mass and evolutionary status of these objects to show that the
highest Li abundances are indeed reached at the red-giant bump for 
low-mass stars and the early-AGB phase for intermediate-mass stars.
The present analysis brings to light the first observational evidence for the
production of Li in the latter group of stars.

\section{Observational Data}

The main properties of the Li-rich giants are gathered in Table 1. 
The lower limit LogN(Li)$>$1.4 was chosen to classify a star as being 
abnormally lithium rich 
\footnote{This corresponds to the value usually used in the 
literature},
though, as we will explain shortly, not all such
stars are truly abnormal.

The parallaxes and associated errors are taken from the Hipparcos catalogue
(\cite{Hipparcos97}) and stellar luminosities are derived from these.  
Of the 28 stars in the sample, five have
no Hipparcos data.  However HDE 233517 has a rotational period of 47.9 d
measured from low-amplitude light variability presumably caused
by spots (\cite{Balachandranetal00}).  Combined with a rotational 
velocity estimated from spectral line 
broadening  of {\it v}sin{\it i} = 17.6 km$^{-1}$ 
\citen{Balachandranetal00} estimate a lower limit to the luminosity 
of 100 L$_{\odot}$.

Effective temperatures are typically taken
from the same source as the Li abundance.  In cases where there are 
multiple estimates, either the mean or the more reliable estimate is chosen.
Temperature errors listed are those
cited by the authors.  No attempt has been made to evaluate them; they
are typically in the range of 100 to 200 K.  
Metallicities are taken from the 
\citen{cayreletal97} or \citen{taylor99} compilations when available.  
Otherwise they are usually taken from the same source as the Li abundance.  
For multiple listings, the values are typically in good agreement.
Most objects have a metallicity close to solar.
About a third of the objects have carbon isotopic ratio measurements.
Errors are given when listed by the authors.
Rotational velocities and information about binarity is taken primarily 
from the CORAVEL compilation of 
\citen{DeMedeirosetal96} and \citen{DeMedeirosetal99}.
The source for the effective temperature, the metallicity, Li abundance,
carbon isotopic ratio and rotational velocity are 
listed alongside the values in Table 1.  As for temperature and metallicity, 
the error in the estimate of the Li abundance is that cited by the authors.  

Accurate luminosities and temperatures allow placement of the stars on the
HR diagram. Masses were then derived from the Geneva-Toulouse evolutionary 
tracks (see \S 4.1). 
Solar metallicity tracks were used for all stars except HD~33798,
HD~39853 and HDE~233517.  For these, mass estimates were interpolated
from solar and [Fe/H]=-0.5 tracks for the metallicity of the star.
Errors on the masses reflect errors in luminosity and temperature.

Towards the end of this study, four more Li-rich giants were identified
by \citen{Strassmeier00} : HD~6665, HD~109703, HD~203136, and
HD~217352.  All have Hipparcos parallaxes, but other than the
\citen{Strassmeier00} study, there is no published information on 
these stars. Strassmeier et al. (2000) used
Tycho B-V colors (\cite{Hipparcos97}) to derive temperatures and
non-LTE curves of growth from \citen{PavlenkoMagazzu96} to derive
Li abundances.  Because the reddening is unknown,
the temperatures of these stars are uncertain.
Strassmeier (private communication) suggested that the canonical reddening
of 0.1 mag per 100pc may be adopted.  Under such an assumption, the 
listed temperatures
of the 4 stars may be too cool by as much as 300 K to 700 K.
The uncertain temperatures result in uncertain Li abundances.
We have therefore not derived the masses of these stars and we have
not included them in Figure 1.
However these stars are discussed briefly in the subsequent sections,
though we caution that the interpretation is uncertain.
A thorough spectroscopic analysis of these stars would be valuable.

The four stars without parallax measurements: HD 19745, HD 25893, HD 95799 and
HD 203251 are not discussed further.

\begin{table*}

\label{tab1}
\rotatebox{90}{
\parbox{210mm}{
\caption{Properties of the Li-rich field giants}
\begin{tabular}{ccccclccllcc}
\hline \\[0.4mm]
\multicolumn{1}{c}{HD}&
\multicolumn{1}{c}{Status}&
\multicolumn{1}{c}{$\pi$ $^a$}&
\multicolumn{1}{c}{$\sigma (\pi)$}&
\multicolumn{1}{c}{V}&
\multicolumn{1}{c}{T$_{\rm eff}$}&
\multicolumn{1}{c}{log(L/L$_{\odot}$) $^b$} &
\multicolumn{1}{c}{M/M$_{\odot}$ $^c$}  &
\multicolumn{1}{c}{[Fe/H]} &
\multicolumn{1}{c}{log N(Li)} &
\multicolumn{1}{c}{$^{12}$C/$^{13}$C}&
\multicolumn{1}{c}{{\it v}sin{\it i}} \\
\multicolumn{1}{c}{number}&
\multicolumn{1}{c}{}&
\multicolumn{1}{c}{(mas)}&
\multicolumn{1}{c}{(mas)}&
\multicolumn{1}{c}{}&
\multicolumn{1}{c}{(K)}&
\multicolumn{1}{c}{}&
\multicolumn{1}{c}{}&
\multicolumn{1}{c}{}& 
\multicolumn{1}{c}{}&
\multicolumn{1}{c}{}&
\multicolumn{1}{c}{(km s$^{-1}$)} \\[1mm]
\hline \\[1mm]
787 & $\sharp$& 5.33&0.87&5.29& 4181$\pm$50(4)& 2.79 & 4.0$\pm$0.5 & +0.03(6)
& 1.80$\pm$0.3(5) & 15(7) & 1.9(9) \\[1mm]
6665 & $\otimes$& 3.53&1.01&8.56& 4500:(17) & 1.65& .. & .. & 2.92:(17) & .. & 10(17)\\[1mm]
9746 &$\star$& 7.77&0.82&6.2&4400$\pm$100(5)& 1.92 & 1.9$\pm$0.3 & $-$0.06(18) &
3.40$\pm$0.2(1) & 28$\pm$4(5) & 8.7(9)\\[1mm]
19745&.. & ..& .. & 9.11 &4990$\pm$100(8) &.. & .. & .. & 4.08$\pm$0.1(8) &
15(7) & 1.0(10)\\[1mm]
21018& $\ominus$ & 2.92 & 0.95 & 6.37 & 5136$\pm$200(2) & 2.52 & 4.7$\pm$0.7 & ..
& 3.35$\pm0.4(2)$ & ..& ..\\[1mm]
25893&..&48.59&1.17&7.13&5300$\pm$200(11) &-0.1& very low & $-$0.10(11) &
1.65$\pm$0.2(11) & .. & 5.1(9)\\[1mm]
30834 &$\sharp$& 5.81&0.82&4.79&4200$\pm$100(5) & 2.85 & 4.5$\pm$0.3 & $-$0.05(6) &
1.80$\pm$0.3(5) & 13(3) & 2.7(9)\\[1mm]
31993 &$\ominus$& 4.2 &1.09&7.48&4500$\pm$200(11) & 1.93 & 2.2$\pm$0.6 &+0.10(11) &
1.40$\pm$0.2(11) & ..& 31.1(11)\\[1mm]
33798 &$\ominus$& 8.94&1.35&6.91&4500$\pm$200(11) & 1.40 & 1.1$\pm$0.5 & $-$0.30(11) &
1.50$\pm$0.2(11) & .. & 29(12)\\[1mm]
39853 &$\sharp$& 4.37&0.72&5.62&3900$\pm$100(14) & 2.81 & 1.5$\pm$0.3 & $-$0.46(14) &
2.80$\pm$0.2(14) & 7(7)& 3.1(9)\\[1mm]
40827 &$\ominus$& 6.92&0.74&6.32& 4575$\pm$100(5) & 1.92 & 2.5$\pm$0.3 &+0.05(18) &
1.60$\pm0.3(5)$ & .. & 1.8(9)\\[1mm]
95799&.. & .. & .. & 7.99 &4800$\pm$200(16) &.. &.. &$-$0.11(16) & 3.22$\pm0.2(16)$ &
10$\pm$7(16)& ..\\[1mm]
108471 &$\ominus$&4.54&0.90&6.36&4970$\pm$100(5) &2.20 & 4.0$\pm$0.3 & $-$0.01(18) &
2.00$\pm$0.3(5) & 25$\pm$7(5) & 4.1(9)\\[1mm]
109703& $\otimes$& 2.82 & 1.08 & 8.70 & 4825:(17) & 1.59& .. & .. & 2.82:(17) & .. & 35.2(17)\\[1mm]
112127 &$\star$&8.09&0.85&6.91&4340$\pm$100(5)& 1.57 & 1.1$\pm$0.2 &+0.09(6) &
2.70$\pm$0.3(5) & 22$\pm$7(19) & 1.6(9) \\[1mm]
116292 &$\ominus$&10.20&0.73&5.36&4870$\pm$100(5) & 1.92 & 3.2$\pm$0.2 & $-$0.15(18) 
& 1.50$\pm$0.3(5) & .. & ..\\[1mm]
120602 &$\ominus$&8.09&0.81&6.00&5000$\pm$100(5) & 1.83& 3.2$\pm$0.2 & $-$0.08(18) &
1.90$\pm$0.3(5) & 16(3)& 5(11)\\[1mm]
121710 &$\sharp$&5.03&0.78&5.02&4100$\pm$100(5)& 2.88 & 3.5$\pm$0.5 & $-$0.27(6) &
1.50$\pm$0.3(5) & .. & 1.3(9)\\[1mm]
126868 &$\ominus$&24.15&1.0&4.81&5500$\pm$100(5) & 1.27 & 2.1$\pm$0.1 & $-$0.07(18)&
2.40$\pm$0.2(5) & .. & 14.4(9)\\[1mm]
146850 &$\sharp$& 3.77&0.85&5.97&4200$\pm$200(13,15)& 2.76 & 4.0$\pm$0.9 & +0.00(20) &
2.00$\pm$0.2(20) & .. &14.4(9)\\[1mm]
148293 &$\star$&11.09& 0.47&5.26&4640$\pm$100(5) & 1.93 & 2.7$\pm$0.4 &+0.08(6) &
2.00$\pm$0.3(5) & 16(3)& 1.2(9)\\[1mm]
183492 &$\star$&11.38& 0.73&5.57&4700$\pm$100(5) &1.75 & 2.5$\pm$0.3 & $-$0.08(6) &
2.00$\pm$0.3(5) & 9(3)& 1.0(9)\\[1mm]
203136& $\otimes$& 4.35& 0.69 & 7.83 & 4983:(17) &1.64  &  .. & .. & 2.23:(17) &  5.4(17)\\[1mm]
203251&.. & .. & ..  &  8.03 &4500$\pm$200(11) &  ..  &    ..        & $-$0.30(11) &
1.40$\pm$0.2(11) & .. & 44.8(9)\\[1mm]
205349& $\ominus$ &2.17& 0.61&6.27&4480$\pm$100(5)& 3.25 & $>$ 5 & .. &
1.90$\pm$0.3(5) & ..& ..\\[1mm]
217352 &$\otimes$ & 5.11& 1& 7.28 & 4569:(17) &1.83  & .. & .. & 2.64:(17) & 35.3(17)\\[1mm]
219025 &$\star$&3.25& 0.81&7.67&4570$\pm$200(13,15)& 2.07 & 2.9$\pm$0.5 & $-$0.10(15) &
3.00$\pm$0.2(15) &..& 25(13)\\[1mm]
233517& $\star$ & ..& ..  &8.41& 4475$\pm$70(1) & 2.0$^d$ & 1.7$\pm$0.2 & $-$0.37(1)
&4.35$\pm$0.1(1) &.. & 17.6(1)\\[1mm]
\hline
\end{tabular}\\
{\small
\begin{tabular}{llll}
$\ominus$ ~~ undergoing standard Li dilution &
$\star$ ~~ at luminosity function bump &
$\sharp$ ~~ at early-AGB phase &
$\otimes$ ~~ uncertain T$_{eff}$ and Li \\
$^a$ ~~from Hipparcos catalogue &
$^b$ ~~derived from Hipparcos data &
$^c$ ~~derived from evolutionary tracks &
$^d$ ~~computed from P$_{rot}$ and {\it v}sin{\it i} \\
\end{tabular}
\begin{tabular}{llll}
(1) Balachandran et al. (2000) &
(2) Barrado y Navascues et al. (1998) &
(3) Berdyugina \& Savanov (1994) &
(4) Blackwell \& Lynas-Gray (1998) \\
(5) Brown et al. (1989) &
(6) Cayrel de Strobel et al. (1997) &
(7) da Silva et al. (1995) &
(8) de la Reza \& da Silva (1995) \\
(9) de Medeiros \& Mayor (1999) &
(10) de Medeiros et al. (1996) &
(11) Fekel \& Balachandran (1993) &
(12) Fekel \& Marschall (1991) \\
(13) Fekel \& Watson (1998) & 
(14) Gratton \& D'Antona (1989) &
(15) Jasniewicz et al. (1999) &
(16) Luck (1994) \\
(17) Strassmeier et al. (2000)&
(18) Taylor (1999) &
(19) Wallerstein \& Sneden (1982) &
(20) This paper \\
\end{tabular}
}
}
}

\end{table*}

\begin{figure*}
\centerline{
\psfig{figure=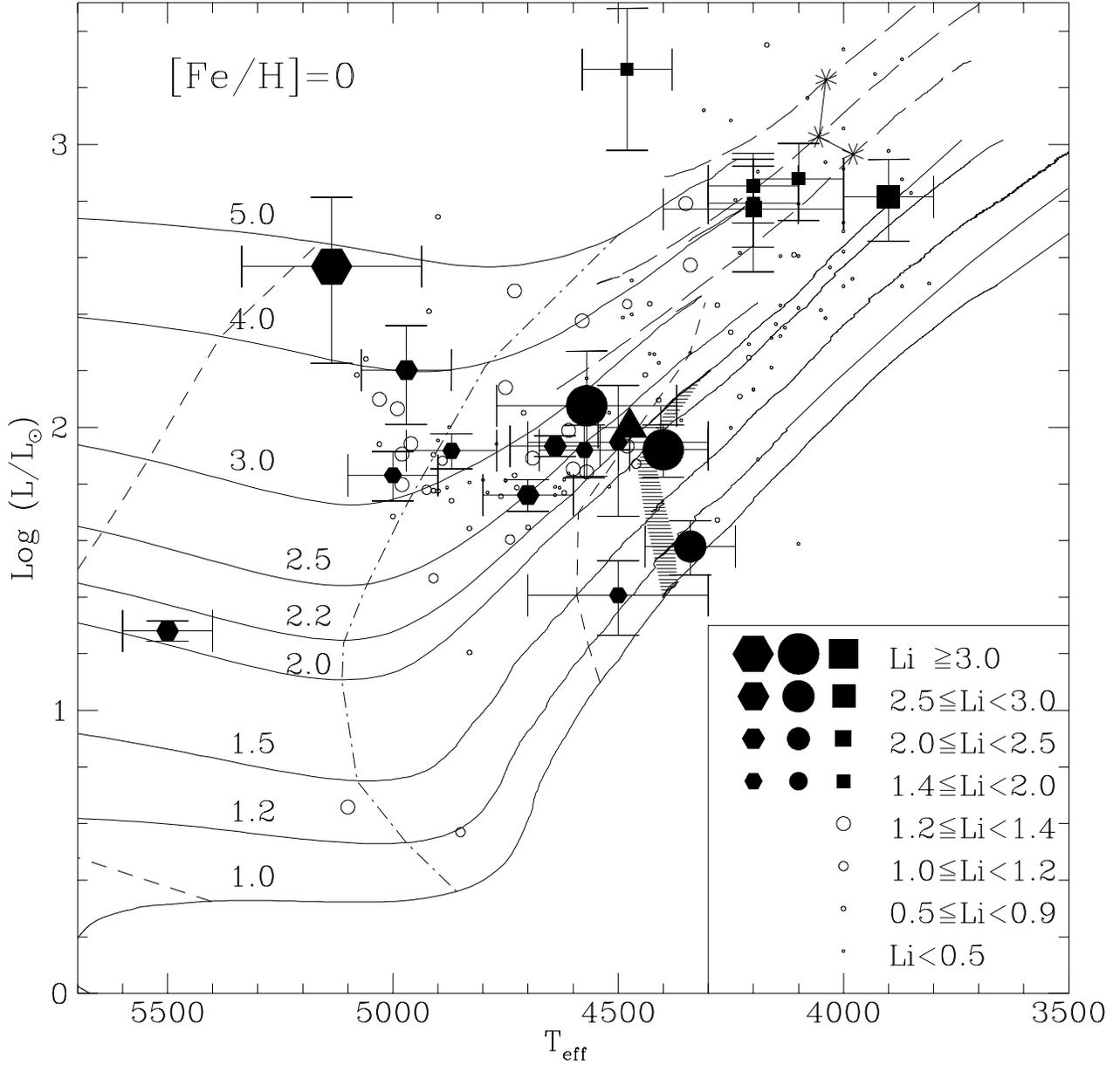,height=18cm}
}
\caption{The HR diagram for the Li-rich giants. 
Evolutionary tracks with [Fe/H]=0 are labelled by their masses.  
Two dashed lines delimit the first dredge-up; the warmer line 
corresponds to the start of Li dilution and the cooler line marks the
deepest penetration of the convective zone.  The dashed-dotted line
indicates the start of the decrease in the carbon isotopic ratio 
due to dilution. 
The shaded region surrounds the location of the RGB bump.
A solid line connects the points (asterisks) where a molecular
weight barrier starts to appear due to the second dredge-up on the
early-AGB.
The filled symbols represent the so-called lithium rich stars 
which are grouped according to their evolutionary state as discussed in
the text: hexagons indicate Li-normal
stars undergoing first dredge-up dilution; circles and triangle 
(for HD~233517 whose luminosity was derived from P$_{rot}$ and {\it v}sin{\it i})
show stars at the RGB bump; and squares indicate the early-AGB stars. 
The open circles are stars with Hipparcos parallaxes and -0.1$\leq$[Fe/H]$\leq$+0.1 
from the Brown et al. (1989) survey.  In all cases,
the size of the symbol indicates the Li abundance of the star as
shown in the legend.
The error bars on luminosity are estimated from uncertainties in the
Hipparcos parallax.  Temperature uncertainties are taken from
the literature (see Table 1).
}
\end{figure*}

\section{Evolutionary Behaviour of Lithium on the HR diagram}
\subsection{The Hertzsprung-Russell diagram} 
To determine the mass and the evolutionary status of the Li-rich 
giants, we have computed new stellar tracks 
with the Geneva-Toulouse evolutionary code with up-to-date input physics (same 
as in \cite{ct99}), but with no diffusion or rotationally induced mixing.
The tracks range in mass from 1.0 to 5.0 M$_{\odot}$ 
and in [Fe/H] from 0.0 and -0.5. 
The models were computed from the pre-main sequence up to the helium flash
for the low-mass stars and up to the early-AGB for the intermediate-mass stars 
which ignite He in a non-degenerate core (Fig. 1). 
For clarity, we show only the advanced evolutionary phases for the [Fe/H]=0.0 tracks.
The subgiant and RGB evolutionary tracks are shown by solid
lines and the early-AGB phase for masses higher than 3 M$_{\odot}$ 
is shown by long dashed lines.
On the HR diagram, two dashed lines show the beginning of Li dilution 
(warmer line) and the end of the first dredge-up (cooler line). 
Between these two, the dashed-dotted line shows the onset of the carbon 
isotopic ratio dilution. 
Because lithium burns at a very low temperature, the deepening convective 
envelope reaches the lithium-free region before it engulfs the $^{13}$C peak
built on the main sequence. 
Also shown in Fig. 1 is the RGB bump indicated by the shaded region.  
Only low-mass stars that have a highly degenerate He core on the RGB, and 
later undergo the He flash, evolve through this phase. In the 
intermediate-mass stars, helium burning in the core
begins before the hydrogen-burning shell reaches the chemical
discontinuity. 
Finally, a solid line connects the points (asterisks) where a 
molecular weight barrier starts to appear due to the second dredge-up 
on the early-AGB (long dashed lines).

In Figure 1, all the Li-rich giants for which we are able to estimate
luminosities are plotted with filled symbols. The four possible
Li-rich giants from Strassmeier et al. (2000) are excluded because
their temperatures and subsequently their Li abundances are uncertain.
They are however discussed in the text.
The open circles are data with -0.1$\leq$[Fe/H]$\leq$+0.1 from the 
large sample of giants studied by Brown et al. (1989). 
Luminosities for these stars are also
based on Hipparcos parallaxes.  In each case, the size 
of the symbol indicates  
the Li abundance of the star as indicated in the legend.  
Error bars on the Li-rich giants show uncertainties in the luminosity
and temperature estimates of these stars.  Luminosity errors are based on
parallax errors and temperature errors are as cited by the authors from whom
the temperature estimate is taken (see Table 1).

At first glance, the Li-rich giants seem to be randomly distributed 
throughout the 
diagram. However, a more careful analysis leads us to divide the sample
into three groups according to their evolutionary state. Each of these
is now discussed separately.

\subsection{Stars undergoing standard first dredge up dilution}

The stars discussed in this section are indicated as filled hexagons
in Figure 1.
Accurate parallax measurements which lead to luminosity and mass
determinations show that the less evolved stars,
HD 21018, HD 108471, HD 116292, HD 120602, HD 126868 and HD 205349 are 
not true Li-rich giants.
With Li abundances larger than the upper limit of logN(Li)=1.4, 
the maximum theoretically allowed for low-mass giants which have completed 
the first dredge-up, these stars have been classified as Li-rich in the 
literature. 
However their position on the HR diagram shows them to be normal.
HD 205349 with a mass greater than 5 M$_{\odot}$ is probably a supergiant
crossing the Hertzsprung gap.
HD 126868 and HD 21018, 2.1M$_{\odot}$ and 4.7 M$_{\odot}$ subgiants
respectively,
have only just begun their Li dilution phase.  HD 21018 is a chromospherically
active binary.  It was discussed in some detail as being abnormally Li-rich
by both \citen{Barradoetal98} and \citen{Mallik99}.  This is
clearly not the case.  At 4M$_{\odot}$, HD 108471 has begun but not 
completed its Li dilution.  Standard models predict a dilution of 
roughly a factor of 7.5 for HD 108471.  
If it left the main sequence with its surface Li intact, its 
measured dilution is about a factor of 20.  With a mass of
3.2M$_{\odot}$, both HD 116292 and HD 120602 
are also currently undergoing the first dredge-up dilution. 
Rather surprisingly, the $^{12}$C/$^{13}$C ratio of HD 120602 has been measured
at 16 (\cite{BerdyuginaSavanov94}).  According to the standard models, this star 
should, at best, 
have only begun to dilute its 
$^{12}$C/$^{13}$C ratio from the solar value of near 90.  It should 
reach a $^{12}$C/$^{13}$C ratio of
near 25 to 30 at the base of the red giant branch when the standard dilution
process has been completed.  In the compilation of $^{12}$C/$^{13}$C values, 
\citen{CharbonnelDias98}
did not find any other stars like HD 120602.   Its
$^{12}$C/$^{13}$C ratio requires closer scrutiny.

Despite their normal Li abundances, HD 108471, HD 120602 and HD 116729
appear to be unusual because a number of stars in their vicinity on the
HR diagram show far smaller Li abundances (Figure 1).  We argue that this 
merely indicates that non-standard Li depletion has occurred in the other
stars during the main sequence phase.  While 3M$_{\odot}$ stars on 
the main sequence are not typically seen to be depleted in Li,
observations have long suggested that the Li-preservation zones in
these stars may not be as deep as predicted by the standard models, i.e.,
once the star evolves off the main sequence and its convective zone deepens,
the Li abundance decreases far more rapidly than standard predictions 
(\cite{ct99} and references therein). 
Unlike these stars, HD 108471, HD 120602 and HD 116729 have Li-preservation 
zone depths that are in agreement with the predictions of the standard models.

HD 203136 from \citen{Strassmeier00} is in close proximity to HD 120602 on 
the HR diagram and has a comparable Li abundance within the errors.  
Standard models predict an abundance of log N(Li)=2.28, essentially
identical with the measured abundance.
HD 109703 from \citen{Strassmeier00} has an unreddened T$_{eff}$ of 4825 K
which places it at a slightly more evolved phase than 
HD 203136.  Its measured Li abundance of log N(Li)=2.82 is much larger than 
the standard model prediction of log N(Li)=1.75. 
Of the four stars from \citen{Strassmeier00}, HD 109703
is the most distance and it may have a reddening as large as E(B-V)=0.3.
Such a large reddening would increase its 
temperature to 5500 K and the near-meteoritic Li abundance that would
be derived from the increased temperature would be compatible with
standard model predictions.  However, given its Galactic latitude of 
b=+65.45$^{\circ}$, it is unlikely that HD 109703 has such a large reddening and
it remains an enigmatic star that requires further scrutiny.

Three stars in our sample show only marginal Li enhancement:
HD 31993, HD 33798, and HD 40827. It remains questionable
whether these should in fact be labelled as Li-rich. According to their
positions on the HR diagram, these stars should have completed their Li
dilution phase.  Their Li abundances, close to the maximum allowed by the
standard model, suggest that they have not undergone any significant depletion
during the main sequence.  Their masses are consistent with this conclusion;
with masses of 2.2 and 2.5 M$_{\odot}$ respectively, HD 31993 and HD 40827 have 
evolved from the hot side of the Li dip and with a mass of 
1.1 M$_{\odot}$ 
HD 33798 has evolved from the Li plateau region between the cool side of the 
dip and the G-dwarf Li decline where Li depletion is minimal.
We suggest that these stars are not Li-rich, but having retained their initial 
Li abundance during their main sequence lifetimes, 
now exhibit standard post-dilution Li values.

\subsection{Stars at the bump}

The stars discussed in this section are shown as filled circles and
a filled triangle in Figure 1.
Four stars with the highest Li abundances, HD 9746, HD 112127, HD 219025 
and HD 233517, occupy a clump on the HR diagram coincident
with the position of the red-giant bump.  Of these HDE 233517 is
the most Li-rich with a super-meteoritic Li abundance of log N(Li)=4.35 
(\cite{Balachandranetal00}).
Two pieces of observational evidence confirm that the large Li 
abundances in these stars is the result of fresh synthesis.  
First, beryllium depletion has been measured in two of the
stars, HD 9746 and HD 112127 (\cite{Castilhoetal99}) confirming 
that the second light element has undergone the expected dilution.
Second, $^6$Li is not detected in HD 9746 and HDE 233517;
although this was not a direct measurement, \citen{Balachandranetal00}
found that consistent Li abundances are only measured from the resonance
and excited lines of Li I when
$^7$Li alone is considered in the spectral synthesis.
The absence of $^6$Li also confirms that the observed Li is not entirely a
remnant of the initial Li with which these stars were formed.
The high Li abundances in these stars are consistent with our
hypothesis that stars may undergo Li production at the red-giant bump
via the \citen{CameronFowler71} mechanism.
Carbon isotopic ratios have been measured for two of these stars: HD 9746 
and HD 112127.  The standard $^{12}$C/$^{13}$C ratios of 28 (\cite{Brownetal89})
and 22 (\cite{WS82}) respectively
are also consisent with our hypothesis that 
Li-production precedes the extra-mixing phase which connects the convective
envelope with the CN-burning region to produce the low, non-standard 
values of $^{12}$C/$^{13}$C which are observed in 96\% of the evolved giants 
(\cite{CharbonnelDias98}). 
Three of the four stars have masses between 1.9 and 2.9 M$_{\odot}$ 
while HD 112127 has a lower mass of 1.1 M$_{\odot}$.

HD 148293 and HD 183492, are also in the region of the red-giant bump.   
These two stars have a Li abundance of log N(Li) = 2.0 and $^{12}$C/$^{13}$C 
ratios 
of 16 and 9 respectively.  Despite their perceived location to the left of 
the red-giant bump on the HR diagram, their low $^{12}$C/$^{13}$C ratios suggest that
they have evolved past the red-giant bump.  We suggest two possible explanations.
Either the temperatures of these stars are lower or the $^{12}$C/$^{13}$C 
ratios are
larger than estimated.  Temperature errors are estimated at $\pm$100 K by
\citen{Brownetal89}.  
The stars are about 200 K from the theoretically estimated position
of the red-giant bump.  No error estimates are provided for the 
$^{12}$C/$^{13}$C measurements.  We are inclined to believe that the 
temperature shift required to place them at the red-giant bump
is not unduly large given the general uncertainties in temperature measurement
and the inhomogeneity in the sources from which the temperatures are gathered.
With this caveat, we suggest that both stars are at the red-giant bump.
According to the hypothesis we outlined in \S2, and the measured 
$^{12}$C/$^{13}$C ratios, HD 148293 and HD 183492 have completed the Li 
production phase and are currently undergoing ``extra-mixing" leading
to their lower $^{12}$C/$^{13}$C ratios. 

While the Li abundances of HD 148293 and HD 183492 are large, they
are not as large as the four extremely Li-rich giants discussed above.
According to our hypothesis, the Li production phase must be very short-lived.
This is based on observations which indicate 
an abrupt change in the behavior of the carbon isotopic 
ratio shortly after the luminosity-function bump
(see \cite{CharbonnelDP00} for a recent review and references therein)
Once the mixing process which connects the convective envelope to the
CN-burning region sets in, 
the envelope material is exposed to a much higher temperature 
than Li can withstand, and a partial or complete destruction 
of the freshly synthesized is expected 
\footnote{In Pop II giants, the decrease of the carbon isotopic ratio 
due to extra-mixing after the bump is clearly 
associated to a drop of the Li abundance 
(\cite{Pilachowskietal93}, \cite{cchar:95}, \cite{Grattonetal2000}.)}.
The current Li abundances of HD 148293 and HD 183492 therefore reflect a 
destruction of Li from their peak values which may have matched 
that of HDE 233517.  

The position of HD 148293 and HD 183492 on the HR diagram, their
Li abundance, and their $^{12}$C/$^{13}$C ratios provide strong constraints on 
the timescale of the extra-mixing process.  
The abundance changes appear to be abrupt both in terms of luminosity 
and of time. Note that the time spent by the star at the bump
corresponds to about 3$\%$ of the time along the ascent of the RGB 
(for a 1.5M$_{\odot}$ of solar metallicity, this corresponds to about 10Myr).
The mixing episode may last only a fraction of the time spent at the 
bump.
Even if we assume that all stars go through the Li-rich phase, the
swiftness of the phenomenon may explain the relative rareness of the 
Li-rich objects.  Once the mixing process reaches the CN-burning
regions, the low $^{12}$C/$^{13}$C ratio remains to tell the tale of the 
mixing episode, while the fleeting Li-rich phase vanishes.

In the absence of significant reddening, the \citen{Strassmeier00} stars 
HD 6665 and HD 217352 appear to be in close proximity to
the luminosity bump and their high Li abundances are compatible with the
other Li-rich giants seen in this region.  These two stars warrant further
spectroscopic study.

The HR diagram provides another striking point which strengthens our hypothesis.
A second clump of Li-rich giants is seen at a higher luminosity and 
we will discuss this in \S4.4.  With the exception of these stars there are 
no Li-rich giants detected at any evolutionary point between the bump and the 
tip of the RGB.  
Yet, for a 1.5M$_{\odot}$ star of solar metallicity for example, 
the duration required to cover the evolutionary phase between the bump
and the tip of the RGB corresponds to about 20$\%$ of the giant's lifetime, 
i.e., about ten times longer than the duration spent on the bump.
If the fresh Li produced at the bump was preserved in spite of the 
extra-mixing which decreases the carbon isotopic ratio, 
or if Li production was a continuous process all along the
RGB (see \S5.4), some Li-rich giants would surely have been 
detected between these two evolutionary points.

\subsection{Stars on the early-AGB}
A second clump of Li-rich giants is seen at higher luminosity 
(filled squares in Fig. 1). 
HD 787, HD 30834, HD 39853, HD 121710 and HD 146850 have relatively high 
masses (most between 3 and 4.5 M$_{\odot}$). As discussed earlier,
these intermediate-mass stars do not undergo a bump phase on the RGB
as their hydrogen-burning shell does not cross the chemical discontinuity 
created by the first dredge-up until core He burning has commenced.
Since the convective envelope is very shallow during core He burning, 
extra-mixing may not be effective in altering the surface abundances 
at this stage.
Once He in the core has been exhausted, the star begins its ascent of the
AGB. The convective envelope deepens and the models show that in the stars
at the Li-rich clump, it has not reached its maximum depth.
Our hypothesis speculates that it is at this phase that an extra-mixing
episode can most easily connect the $^3$He-rich envelope with interior
temperatures hot enough to fuel the Cameron-Fowler (1971) process, and
this is precisely what the observations suggest occurs.
However as soon as the convective envelope reaches regions which have been 
modified by nuclear processing, a new molecular weight barrier is
established, and the extra-mixing episode should stop.  
We point to several observational aspects which are consistent with
our hypothesis.  First, Li-rich early-AGB stars have not been detected
at higher luminosities where the molecular weight gradient would be
re-established (above the asterisked points in Figure 1).
 Second, although improved statistics would be valuable, 
we note that the Li abundances in these stars are lower than 
in the low mass RGB bump stars consistent with their smaller $^3$He reservoir.
Finally, the Li-rich early-AGB stars 
appear to be confined to masses $<$ 4.5 M$_{\odot}$; in stars of higher mass,
the hydrogen-burning shell is extinguished by the deepening convective zone
and therefore $^7$Li production will not be predicted to occur according
to our hypothesis.

Other surface abundance changes are expected to be induced by this event 
for the elements processed in the hydrogen-burning shell 
and these remain to be studied. 
In particular, the $^{12}$C/$^{13}$C ratio may be modified.
This value has been 
determined in three stars. In HD 787 and HD 30834 (which have masses of 
4 and 4.5 M$_{\odot}$ respectively), the $^{12}$C/$^{13}$C ratio
is only marginally different from 
the values predicted by standard dredge-up for this mass range.
This confirms that these stars did not undergo extra mixing on the RGB.
On the other hand, HD 39853 has a significantly lower carbon isotopic ratio, 
and is also the most Li-rich object of this group. However at 
1.5 M$_{\odot}$, it is also the only low-mass, relatively metal-poor
([Fe/H]=-0.46) star in this group and it must have gone through the RGB 
bump phase. 
Its low carbon isotopic ratio of 7 is in agreement with the values
observed in low mass RGB stars with the same metallicity which have
already gone through the bump (Charbonnel et al. 1998), and is certainly
a signature of the first extra-mixing episode.
We suggest that HD 39853 is actually a unique object in the sample 
since this star is undergoing Li production for the second time 
in its evolution. 

\subsection{Li-rich giants in open and globular clusters}

Three Li-rich giants have recently been discovered in one open and two
globular clusters, namely Berkeley 21 (\cite{HillPasquini99}), 
M3 (\cite{Kraftetal99}) and NGC 362 (\cite{Smithetal99}). 

Berkeley 21 is a relatively metal-poor open cluster 
([Fe/H]=-0.54) with
an age of about 2.2 to 2.5 Gyr (\cite{Tosietal98}) and a corresponding
turnoff mass of about 1.4M$_{\odot}$.
Pasquini \& Hill (1999) observed three giants in Berkeley 21.  All three are
located in close proximity to each other on the HR diagram and only 
one was found to be Li-rich with a near-meteoritic Li abundance; the other
two have Li upper limits of log N(Li) $<$ 0.5.  The He-burning
clump is evident on the color-magnitude diagram of Berkeley 21, and
the observed giants are about 0.8 mag brighter than the clump
(\cite{Tosietal98}; Tosi, private communication).  Theoretical models 
(\cite{cmms96grille6}) place the bump at about 0.5 mag above the 
red-giant clump.  Therefore all 3 giants are consistent with being 
low-mass stars at or evolved slightly past the luminosity bump.  
The Li-rich star has a low carbon isotopic ratio (Hill, private
communication).  We suggest that this star is either on the RGB 
in which case it must have had a much higher peak Li abundance which
has since been depleted, or it has evolved off the core-He burning clump
and is undergoing a second Li-rich episode, similar to HD 39853.

The two Li-rich giants in the globular clusters M3 and NGC 362 are not 
as easily explained.  The stars are bright, putting them either near the 
tip of the RGB or in the early-AGB phase. 
However, unlike the early-AGB stars discussed in \S4.4, in these low-mass 
stars with very low metallicity the second dredge-up is particularly shallow. 
Therefore, mixing would have to be extremely efficient to bridge
the gap between the base of the convective envelope and the hydrogen-
burning shell.
At this stage, we shall not attempt to compartmentalize the globular
cluster stars with the two classes as that is not easily done yet.  
The discovery of additional Li-rich giants in
globular clusters should provide further clues to the nature of 
these stars.

\section{Li-Enrichment Mechanisms}
Our study does not endorse any triggering mechanism for the extra-mixing.  
However, in the light of the data we have collected in this study we offer 
some insights into the various mechanisms suggested in the literature. 
We caution that in the final analysis, these enrichment mechanisms remain 
inconclusive.

\subsection{Rotation}

Rotation leading to meridional circulation, or differential
rotation leading to turbulence have been regarded as the prime mechanisms
for extra-mixing (\cite{sm:79}, \cite{cchar:95}). 
Fekel \& Balachandran (1993), for example, suggested
that angular momentum may be dredged-up from the stellar interior along
with $^7$Li-rich material during the process in which a Li-rich giant
is created.  We looked for a possible correlation between Li
enrichment and rapid rotation in the data.
Of the stars near the RGB bump, HD 219025 and HD 233517 have large rotational
velocities of 25 and 18 km s$^{-1}$ respectively.  However the 
Li-rich giants HD 9746 and HD 112127 have smaller rotational velocities of 
8.7 and 1.2 km s$^{-1}$ respectively and two stars which we claim have 
undergone only normal depletion, HD 31993 and HD 33798, also have large 
rotational velocities of 31 and 29 km s$^{-1}$ respectively. 
Although there is no clear correlation between 
Li abundance and rotational velocity in the stars near the RGB bump, 
perhaps none is expected if Li enrichment and angular momentum dredge-up
proceed at different timescales.  For example, the dredge-up of 
angular momentum may not be hindered by the mean molecular weight gradient
as the material mixing is.  Without clear insight into the associated physical
processes further conclusions may not be drawn.

\subsection{Mass Loss}
In a series of papers, de la Reza and collaborators (\cite{DeLaRezaetal96}, 
\cite{DeLaRezaetal97})
have suggested that all low-mass stars go through one, possibly
several, Li-enrichment episodes each of which is accompanied by a 
mass-loss event. This hypothesis was based on their finding that a
majority of the Li-rich giants have far-infrared color excesses as
measured from IRAS fluxes.  The reader is referred to the original
papers for further details.  Although no triggering
mechanism for the Li-enrichment process was suggested, the mass-loss
scenario, if real, may provide leads for future investigations.  In this
context, we make the following observations.  De la Reza et al. (1996) 
noted that several Li-rich giants had 60-25 micron excesses (see their
Figure 2).  Among these are HD 108471 and HD 120602 which are undergoing
standard first dredge-up and HD 31993 which has completed standard first
dredge-up; these are not Li-rich.  Of the early-AGB stars, HD 787
and HD 39853 and HD 121710 show no evidence of a dust shell, while
HD 30834 and HD 146850 do.  Fekel \& Watson (1998) and Jasniewicz et al. 
(1999) independently obtained Li abundances in stars known to have far-infrared
excesses and found none to be Li-rich.  Although some Li-rich stars 
certainly have a far-infrared excess, and HDE~233517 is an example, there is 
no one-to-one correlation as hypothesized by de la Reza et al.

\subsection{Planet Accretion}
Several studies (\cite{Alexander67}, \cite{Brownetal89} and most
recently, and in most detail, \cite{SiessLivio99}) 
studied the accretion of
a planet or a brown dwarf by an RGB star to produce a Li-rich giant.
The reader is referred to these papers for details of the accretion 
process.  There are several inconsistencies that make it unlikely
that the observed Li-rich giants are made by this process.
First, the accretion scenario encounters some problems in reproducing the 
very high log N(Li) values (i.e., $\geq$ 2.8), for which the authors have to invoke
anyway an extra-mixing process (which could be triggered by the instabilities
related to the accretion).
Also, the accreted matter would result in a simultaneous enhancement of $^6$Li,
$^7$Li and Be. Castilho et al. (1999) have shown that four Li-rich giants
show no evidence of Be-enrichment.  And Balachandran et al. (2000)
have shown that the Li abundances derived from the resonance and excited 
Li I features in the Li-rich giants HDE~233517 and HD~9746 are matched
only when $^6$Li is not included in the synthesis, i.e., $^6$Li is depleted
in these stars.
In addition to these chemical disagreements, planet
accretion should be more easily achieved when the star is further
up the RGB and has a larger radius, and the Li-enrichment at this
phase should be larger as the convective zone is less deep. 
Yet, no Li-rich stars are seen between the RGB bump and the RGB tip.
All of these factors combine to suggest that the accretion of a planet 
leading to the formation of a Li-rich giant may be very rare indeed.

\subsection{Deep Circulation Model}

There is little disagreement among the various studies, that if $^7$Li is 
indeed manufactured in the Li-rich giants, it must occur by the Cameron-Fowler
(1971) process.  Sackmann \& Boothroyd (1999) have recently suggested 
an ad-hoc two stream ``conveyor-belt" circulation model, for circulating
the $^3$He-rich envelope of the red-giant to the H-shell burning temperatures
to achieve $^7$Be production.
In their model, the
depth of the extra-mixed region is related to a parametrized temperature 
difference up to the bottom of the hydrogen-burning shell. 
They demonstrate that certain assumptions, which depend critically on
the mixing speeds and geometry of the ascending and descending streams,
as well as the episodic nature of the mixing process,
could lead to Li creation along the RGB with abundances in excess of
log N(Li)=4.
In particular, high Li enrichment was obtained when mixing was
continuous along the RGB and this simultaneously led
to a smooth decrease of surface $^{12}$C/$^{13}$C ratio; the observed
low values of surface $^{12}$C/$^{13}$C are only reached at the tip of
the RGB.

In contrast to their predictions that Li-rich giants would be produced all
along the RGB, we have noted finding only a single clump at the 
luminosity bump phase of the low-mass stars and a second clump in the 
early-AGB phase of low and intermediate-mass stars.  Furthermore, 
the predicted behavior of the $^{12}$C/$^{13}$C ratio 
in the models differs from the observations, which show that the 
$^{12}$C/$^{13}$C drops sharply just beyond the bump
and then stays constant at a value that depends on the 
stellar mass and metallicity (see Charbonnel et al. 1998). 
Sackmann \& Boothroyd's ``evolving" models are thus not sustained by the 
observations.
Rather, the observed behavior is better simulated by the single short 
lived mixing episode of \citen{SackmannBoothroyd99} under the assumption of a
particularly fast stream.   Of course, the driving force behind the stream 
remains a mystery.

\section{Conclusions}
On the basis of their position on the HR diagram, we are able to separate the
so-called Li-rich giants into three different groups. 
The first group consists of normal stars which have only recently 
started lithium dilution, or have normal post-dilution lithium 
abundances, and are thus mis-labelled as Li-rich.
The second group contains low-mass stars at the luminosity bump on the RGB.
The third group contains 
low and intermediate-mass stars in the early-AGB phase.
The second and third groups support our hypothesis that Li-rich stars
are formed by an extra-mixing process which is effective when the 
convective zone is in close proximity to the hydrogen-burning shell and
when the two regions are not separated by a strong gradient of molecular
weight. 
Li production will be followed by a decrease in the $^{12}$C/$^{13}$C
ratio as the material mixes to deeper layers.  When this occurs the
freshly synthesized Li will be steadily destroyed.  Other elements 
processed in the hydrogen-burning shell may
be affected and this remains to be studied. As the Li production
phase is short and these stars have only a moderate mass loss rate, they
are not expected to contribute significantly to the Li enrichment of the ISM. 

\begin{acknowledgements}
SCB would like to thank John Carr and Frank Fekel for useful comments on the 
paper and is pleased to acknowledge support from NSF grants 
AST-9618335 and AST-9819870. 
CC acknowledges support from the french GdR Galaxies and the ASPS. 
We thank Vanessa Hill and Monica Tosi for fruitful discussions.
We used the SIMBAD data base operated at the CDS (Strasbourg, France).
\end{acknowledgements}

\appendix

\bibliographystyle{MS9657_bst}
\bibliography{MS9657_bib}

\begin{thebibliography}{58}

\bibitem[\protect\astroncite{{Alexander}}{1967}]{Alexander67}
{Alexander} J.B., 1967, The Observatory 87, 238

\bibitem[\protect\astroncite{{Balachandran} et~al.}{2000}]{Balachandranetal00}
{Balachandran} S.C., {Henry} G., {Fekel} F.C., {Uitebroek} H., 2000, \apj, in
  press

\bibitem[\protect\astroncite{{Barrado y Navascues}
  et~al.}{1998}]{Barradoetal98}
{Barrado y Navascues} D., {de Castro} E., {Fernandez- Figueroa} M.J., {Cornide}
  M., {Garcia Lopez} R.J., 1998, \aap 337, 739

\bibitem[\protect\astroncite{{Berdyugina} \&
  {Savanov}}{1994}]{BerdyuginaSavanov94}
{Berdyugina} S.V., {Savanov} I., 1994, Astronomy Letters 20, 639

\bibitem[\protect\astroncite{{Brown} et~al.}{1989}]{Brownetal89}
{Brown} J.A., {Sneden} C., {Lambert} D.L., {Dutchover} E. J., 1989, \apjs 71,
  293

\bibitem[\protect\astroncite{{Cameron} \& {Fowler}}{1971}]{CameronFowler71}
{Cameron} A.G.W., {Fowler} W.A., 1971, \apj 164, 111

\bibitem[\protect\astroncite{{Carretta} et~al.}{1998}]{Carrettaetal98}
{Carretta} E., {Gratton} R.G., {Sneden} C., {Bragaglia} A., 1998, in Galaxy
  Evolution: Connecting the Distant Universe with the Local Fossil Record.
  Organized by Paris-Meudon Observatory from 21-25 September, 1998 in Meudon
  (France). p.~E17

\bibitem[\protect\astroncite{{Castilho} et~al.}{1999}]{Castilhoetal99}
{Castilho} B.V., {Spite} F., {Barbuy} B. et~al., 1999, \aap 345, 249

\bibitem[\protect\astroncite{{Cayrel de Strobel} et~al.}{1997}]{cayreletal97}
{Cayrel de Strobel} G., {Soubiran} C., {Friel} E.D., {Ralite} N., {Francois}
  P., 1997, \aaps 124, 299

\bibitem[\protect\astroncite{{Charbonnel}}{1994}]{cc94}
{Charbonnel} C., 1994, \aap 282, 811

\bibitem[\protect\astroncite{Charbonnel}{1995}]{cchar:95}
Charbonnel C., 1995, ApJL 453, L41

\bibitem[\protect\astroncite{Charbonnel et~al.}{1998}]{cbw:98}
Charbonnel C., Brown J.A., Wallerstein G., 1998, A\&A 332, 204

\bibitem[\protect\astroncite{Charbonnel et~al.}{2000}]{CharbonnelDP00}
Charbonnel C., Deliyannis C., Pinsonneault M., 2000, in The light elements and
  their evolution, no. 198 in IAU Symp. Kluwer, Dordrecht, p.~0

\bibitem[\protect\astroncite{{Charbonnel} \& {Do
  Nascimento}}{1998}]{CharbonnelDias98}
{Charbonnel} C., {Do Nascimento} J.~D. J., 1998, \aap 336, 915

\bibitem[\protect\astroncite{{Charbonnel} et~al.}{1996}]{cmms96grille6}
{Charbonnel} C., {Meynet} G., {Maeder} A., {Schaerer} D., 1996, \aaps 115, 339+

\bibitem[\protect\astroncite{{Charbonnel} \& {Talon}}{1999}]{ct99}
{Charbonnel} C., {Talon} S., 1999, \aap 351, 635

\bibitem[\protect\astroncite{Crane}{1994}]{Crane94}
Crane P., 1994, in P. Crane (ed.), The Light Element Abundances, ESO
  Astrophysics Symposia. Springer, Berlin, p.~0

\bibitem[\protect\astroncite{{Da~Costa}}{1998}]{daco:98}
{Da~Costa} G.S., 1998, in T.~R. Bedding, A.~J. Booth, and J. Bavis (eds.),
  Fundamental Stellar Properties: the interaction between observation and
  theory, no. 189 in IAU Symp. Kluwer, Dordrecht, p.~193

\bibitem[\protect\astroncite{{de la Reza} et~al.}{1997}]{DeLaRezaetal97}
{de la Reza} R., {Drake} N.A., {da Silva} L., {Torres} C.A.O., {Martin} E.L.,
  1997, \apjl 482, L77

\bibitem[\protect\astroncite{{de la Reza} \& {da
  Silva}}{1995}]{DeLaRezaDaSilva95}
{de la Reza} R., {da Silva} L., 1995, \apj 439, 917

\bibitem[\protect\astroncite{{de la Reza} et~al.}{1996}]{DeLaRezaetal96}
{de la Reza} R., {Drake} N.A., {da Silva} L., 1996, \apjl 456, L115

\bibitem[\protect\astroncite{{De Medeiros} \& {Mayor}}{1999}]{DeMedeirosetal99}
{De Medeiros} J.R., {Mayor} M., 1999, \aaps 139, 433

\bibitem[\protect\astroncite{{De Medeiros} et~al.}{1996}]{DeMedeirosetal96}
{De Medeiros} J.R., {Melo} C.H.F., {Mayor} M., 1996, \aap 309, 465

\bibitem[\protect\astroncite{{Fekel} \&
  {Balachandran}}{1993}]{FekelBalachandran93}
{Fekel} F.C., {Balachandran} S., 1993, \apj 403, 708

\bibitem[\protect\astroncite{{Forestini} \&
  {Charbonnel}}{1997}]{ForestiniCharbonnel97}
{Forestini} M., {Charbonnel} C., 1997, \aaps 123, 241

\bibitem[\protect\astroncite{{Gilroy}}{1989}]{Gilroy89}
{Gilroy} K.K., 1989, \apj 347, 835

\bibitem[\protect\astroncite{{Gilroy} \& {Brown}}{1991}]{GilroyBrown91}
{Gilroy} K.K., {Brown} J.A., 1991, \apj 371, 578

\bibitem[\protect\astroncite{Gratton \& D'Antona}{1989}]{GrattonDAntona89}
Gratton R., D'Antona F., 1989, \aap 215, 66

\bibitem[\protect\astroncite{{Gratton} et~al.}{2000}]{Grattonetal2000}
{Gratton} R.G., {Sneden} C., {Carretta} E., {Bragaglia} A., 2000, \aap 354, 169

\bibitem[\protect\astroncite{{Hill} \& {Pasquini}}{1999}]{HillPasquini99}
{Hill} V., {Pasquini} L., 1999, \aap 348, L21

\bibitem[\protect\astroncite{Hogan}{1995}]{hog:95}
Hogan C.J., 1995, ApJL 441, L17

\bibitem[\protect\astroncite{Kraft}{1994}]{kra:94}
Kraft R.P., 1994, PASP 106, 553

\bibitem[\protect\astroncite{{Kraft} et~al.}{1999}]{Kraftetal99}
{Kraft} R.P., {Peterson} R.C., {Guhathakurta} P. et~al., 1999, \apjl 518, L53

\bibitem[\protect\astroncite{{Lambert} et~al.}{1980}]{Lambertetal80}
{Lambert} D.L., {Dominy} J.F., {Sivertsen} S., 1980, \apj 235, 114

\bibitem[\protect\astroncite{{Mallik}}{1999}]{Mallik99}
{Mallik} S.V., 1999, \aap 352, 495

\bibitem[\protect\astroncite{{Pallavicini} et~al.}{1990}]{Pallavicinietal90}
{Pallavicini} R., {Randich} S., {Giampapa} M., {Cutispoto} G., 1990, The
  Messenger 62, 51

\bibitem[\protect\astroncite{{Pavlenko} \& {Magazzu}}{1996}]{PavlenkoMagazzu96}
{Pavlenko} Y.V., {Magazzu} A., 1996, \aap 311, 961

\bibitem[\protect\astroncite{{Perryman} et~al.}{1997}]{Hipparcos97}
{Perryman} M.A.C., {Lindegren} L., {Kovalevsky} J. et~al., 1997, \aap 323, L49

\bibitem[\protect\astroncite{{Pilachowski} et~al.}{1997}]{Pilachowskietal97}
{Pilachowski} C., {Sneden} C., {Hinkle} K., {Joyce} R., 1997, \aj 114, 819

\bibitem[\protect\astroncite{{Pilachowski} et~al.}{1990}]{Pilachowskietal90}
{Pilachowski} C.A., {Hudek} D., {Sneden} C., 1990, \aj 99, 1225

\bibitem[\protect\astroncite{{Pilachowski} et~al.}{1993}]{Pilachowskietal93}
{Pilachowski} C.A., {Sneden} C., {Booth} J., 1993, \apj 407, 699

\bibitem[\protect\astroncite{{Plez} et~al.}{1993}]{Plezetal93}
{Plez} B., {Smith} V.V., {Lambert} D.L., 1993, \apj 418, 812

\bibitem[\protect\astroncite{{Rood} et~al.}{1984}]{rbw84}
{Rood} R.T., {Bania} T.M., {Wilson} T.L., 1984, \apj 280, 629

\bibitem[\protect\astroncite{{Sackmann} \&
  {Boothroyd}}{1992}]{SackmannBoothroyd92}
{Sackmann} I.J., {Boothroyd} A.I., 1992, \apj 392, L71

\bibitem[\protect\astroncite{{Sackmann} \&
  {Boothroyd}}{1999}]{SackmannBoothroyd99}
{Sackmann} I.J., {Boothroyd} A.I., 1999, \apj 510, 217

\bibitem[\protect\astroncite{{Shetrone} et~al.}{1993}]{Shetroneetal93}
{Shetrone} M.D., {Sneden} C., {Pilachowski} C.A., 1993, \pasp 105, 337

\bibitem[\protect\astroncite{{Siess} \& {Livio}}{1999}]{SiessLivio99}
{Siess} L., {Livio} M., 1999, \mnras 308, 1133

\bibitem[\protect\astroncite{{Smith} \& {Lambert}}{1989}]{SmithLambert89}
{Smith} V., {Lambert} D.L., 1989, \apj 345, L75

\bibitem[\protect\astroncite{{Smith} \& {Lambert}}{1990}]{SmithLambert90}
{Smith} V., {Lambert} D.L., 1990, \apj 361, L69

\bibitem[\protect\astroncite{{Smith} et~al.}{1995}]{Smithetal95}
{Smith} V., {Plez} B., {Lambert} D.L., {Lubowich} D.A., 1995, \apj 441, 735

\bibitem[\protect\astroncite{{Smith} et~al.}{1999}]{Smithetal99}
{Smith} V.V., {Shetrone} M.D., {Keane} M.J., 1999, \apjl 516, L73

\bibitem[\protect\astroncite{{Sneden} et~al.}{1986}]{Snedenetal86}
{Sneden} C., {Pilachowski} C.A., {Vandenberg} D.A., 1986, \apj 311, 826

\bibitem[\protect\astroncite{{Strassmeier} et~al.}{2000}]{Strassmeier00}
{Strassmeier} K., {Washuettl} A., {Granzer} T., {Scheck} M., {Weber} M., 2000,
  \aaps 142, 275

\bibitem[\protect\astroncite{Sweigart \& Mengel}{1979}]{sm:79}
Sweigart A.V., Mengel K.G., 1979, ApJ 229, 624

\bibitem[\protect\astroncite{{Taylor}}{1999}]{taylor99}
{Taylor} B.J., 1999, \aaps 139, 63

\bibitem[\protect\astroncite{{Tosi} et~al.}{1998}]{Tosietal98}
{Tosi} M., {Pulone} L., {Marconi} G., {Bragaglia} A., 1998, \mnras 299, 834

\bibitem[\protect\astroncite{{Wallerstein} \& {Sneden}}{1982}]{WS82}
{Wallerstein} G., {Sneden} C., 1982, \apj 255, 577

\bibitem[\protect\astroncite{Weiss et~al.}{2000}]{weissDC00}
Weiss A., Denissenkov P.A., C.Charbonnel, 2000, A\&A, accepted 000, 000

\end{thebibliography}

\end{document}